\documentclass[aps,prl,twocolumn,groupedaddress,showpacs]{revtex4}
\usepackage{bm}
\usepackage{graphicx}
\usepackage{color}

\begin{document}

\title{Direct Measurement of the Out-of-Plane Spin Texture in the Dirac Cone Surface State of a Topological Insulator}
\author{S. Souma$^1$, K. Kosaka,$^2$ T. Sato,$^2$ M. Komatsu,$^2$ A. Takayama,$^2$ T. Takahashi,$^{1,2}$ M. Kriener,$^3$ Kouji Segawa,$^3$ and Yoichi Ando$^3$}
\affiliation{$^1$WPI Research Center, Advanced Institute for Materials Research, 
Tohoku University, Sendai 980-8577, Japan}
\affiliation{$^2$Department of Physics, Tohoku University, Sendai 980-8578, Japan}
\affiliation{$^3$Institute of Scientific and Industrial Research, Osaka University, Ibaraki, Osaka 567-0047, Japan}

\date{\today}

\begin{abstract}
We have performed spin- and angle-resolved photoemission spectroscopy of Bi$_2$Te$_3$ and present the first direct evidence for the existence of the out-of-plane spin component on the surface state of a topological insulator.  We found that the magnitude of the out-of-plane spin polarization on a hexagonally deformed Fermi surface of Bi$_2$Te$_3$ reaches maximally 25\% of the in-plane counterpart while such a sizable out-of-plane spin component does not exist in the more circular Fermi surface of TlBiSe$_2$, indicating that the hexagonal deformation of the Fermi surface is responsible for the deviation from the ideal helical spin texture.  The observed out-of-plane polarization is much smaller than that expected from existing theory, suggesting that an additional ingredient is necessary for correctly understanding the surface spin polarization in  Bi$_2$Te$_3$.
\end{abstract}

\maketitle

   The topological insulators (TIs) materialize a new state of quantum matter where an unusual gapless metallic state appears at the edge or the surface of a band insulator due to a topological principle.  The surface state (SS) of three-dimensional TIs is characterized by a Dirac-cone dispersion which has been shown to have a helical spin structure where the spin vector points parallel to the surface and perpendicular to the momentum {\bf k}, as shown in Fig. 1(a) \cite{HasanRMP, HasanBi2Se3}.  Because of this helicity in the spin direction and the protection by the time-reversal symmetry, the Dirac fermions in the TIs are immune to the backward scattering \cite{YazdaniSTM, YazdaniSb} and are not very sensitive to nonmagnetic impurities or disorder.  This peculiar situation provides a platform for novel topological phenomena such as the emergence of Majorana fermions in the proximity-induced superconducting state, and indeed many theoretical models or experimental results were developed or interpreted relying essentially on this simple helical spin structure \cite{HasanRMP, YazdaniSb, YazdaniSTM}.  However, it is unclear at the moment to what extent such a simple spin texture is adequate.  In fact, recent theoretical studies predicted that when the Fermi surface (FS) of the surface Dirac state is hexagonally deformed, the spin structure starts to obtain a finite out-of-plane (OP) component \cite{AlpichshevPRL, FuPRL, SCZhangPRB, Louie}.  It was also predicted \cite{FuPRL, Louie} that such a component can be as large as the in-plane (IP) counterpart in Bi$_2$Te$_3$ which shows the strongest hexagonal FS warping among known TIs \cite{AlpichshevPRL, ShenBi2Te3}.  The deviation from the simple helical spin texture would be a key to better understanding the peculiar SS of the TIs, whereas an experimental verification has not yet been successfully made.
   
    In this Letter, we demonstrate that such an OP spin component is indeed present in Bi$_2$Te$_3$, by determining for the first time the detailed IP and OP spin texture at various {\bf k} locations of the Dirac-cone SS using the low-energy, spin- and angle-resolved photoemission spectroscopy (SR-ARPES) \cite{SoumaRSIspin}.  We also elucidated that the magnitude of the OP spin polarization is related to the strength of the hexagonal warping of the FS.  We discuss the present result in relation to theoretical proposals as well as its implications for potential applications.
    
     High-quality single-crystals of Bi$_2$Te$_3$ were grown by melting nearly stoichiometric mixtures of high-purity (99.9999\% pure) elemental shots of Bi and Te in a quartz glass tube at 850$^{\circ}$C for 48 h and slowly cooling to 500$^{\circ}$C over 48 h. To tune the chemical potential to be located close to the bottom of the conduction band (CB), which was crucial for the success of this work, we used a slightly Bi-rich composition of Bi:Te = 0.34:0.66. ARPES measurements were performed by using an SR-ARPES spectrometer at Tohoku University \cite{SoumaRSIspin, SoumaRSIXe}.  We used the Xe I photons (8.437 eV) to excite photoelectrons.  Samples were cleaved {\it in-situ} along the (111) crystal plane in an ultrahigh vacuum of 5$\times$10$^{-11}$ Torr. The energy resolutions during the SR- and regular ARPES measurements were 40 and 6 meV, respectively.  We used the Sherman function value of 0.07 to obtain the SR-ARPES data.
          
Figure 1(b) shows the FS mapping of an $n$-type Bi$_2$Te$_3$ sample.  We clearly find a FS showing a strong hexagonal deformation as in previous reports \cite{AlpichshevPRL,ShenBi2Te3}.  This snowflakelike FS originates from an electronlike surface band [Fig. 1(c)] corresponding to the upper branch of the Dirac cone. The Dirac point is situated at the binding energy $E_{\rm B}$ of 0.25 eV, suggesting that the bottom of the bulk CB is located at slightly above the Fermi level ({\it cf.} Ref. \cite{ShenBi2Te3}); in other words, the chemical potential ($\mu$) lies slightly below the bottom of the bulk CB. This is consistent with the slightly $n$-type character of the crystal indicated by the Seebeck coefficient and is suited for examining the OP spin component because of the large hexagonal warping of the FS at this energy position \cite{ShenBi2Te3}.  It should be emphasized here that the absence of bulk CB electrons below $E_{\rm F}$ makes it possible to quantitatively determine the net spin polarization solely from the surface band by avoiding the contamination from the bulk band.

\begin{figure}[t]
\includegraphics[width=3.4in]{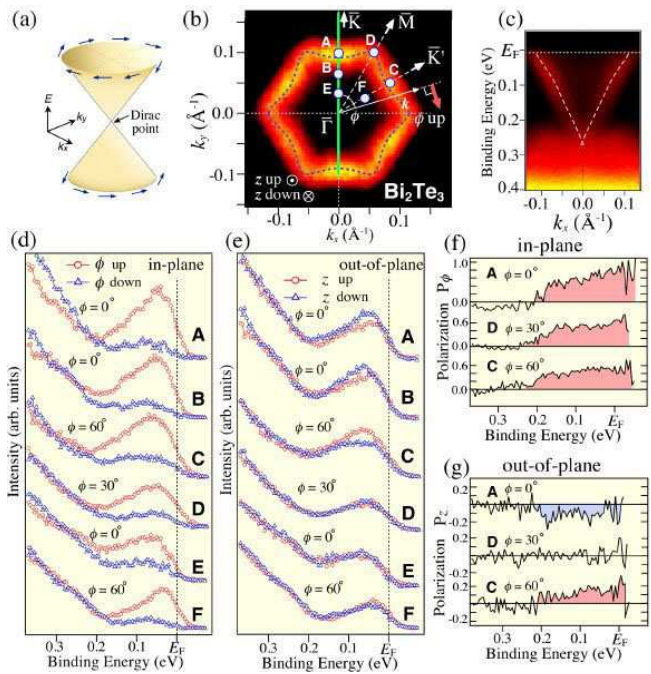}
\caption{(Color online) (a) Schematic picture of the Dirac-cone SS with a simple helical spin texture.  (b) ARPES intensity at $E_{\rm F}$ of an $n$-type Bi$_2$Te$_3$ crystal around the $\bar{\Gamma}$ point plotted as a function of two-dimensional wave vector {\bf k}, highlighting the snowflakelike FS.  The ARPES intensity is integrated within $\pm$20 meV with respect to $E_{\rm F}$ and folded by taking into account the crystal symmetry.  Definition of the FS angle $\phi$ and the IP up-spin vector ($\phi$ up) are also indicated. (c) Near-$E_{\rm F}$ ARPES intensity measured along the $\bar{\Gamma}$$\bar{M}$ cut.  (d),(e)  SR-EDCs for the IP ($\phi$) and OP ($z$) components, respectively, measured at various {\bf k} points indicated by open circles in (b).  (f),(g) IP and OP spin polarizations $P_{\phi}$ and $P_z$ at points A, D, and C.
}
\end{figure}

Figure 1(d) shows spin-resolved energy distribution curves (SR-EDCs) of Bi$_2$Te$_3$ for the IP spin component measured at various {\bf k} locations around the FS indicated in Fig. 1(b).   We define the IP spin polarization vector to point perpendicular to measured {\bf k}  [see, e.g., the thick (red) arrow in Fig. 1(b)] in which the ``up spin'' points to the clockwise direction.  A peak near the Fermi level ($E_{\rm F}$) in the up-spin EDCs originates from the upper branch of the Dirac cone, while a shoulder feature at $E_{\rm B}$ $>$ 0.2 eV is a tail of the bulk valence band.  In contrast to the small differences between the up- and down-spin EDCs for the shoulder feature at $E_{\rm B}$ $>$ 0.2 eV, the difference for the near-$E_{\rm F}$ peak is gigantic: it is dominated by the up-spin component, indicating the basically spin helical nature of the SS where the IP spin vector points to the clockwise direction like in Fig. 1(a).

The important question is whether there is a finite OP spin component perpendicular to the sample surface, $S_z$. We can directly address this issue with our SR-spectrometer, which allows us to simultaneously measure the OP component with the IP component, and Fig. 1(e) shows the SR-EDCs for the OP spin component. If there is no OP spin polarization, one should observe that the EDCs for $S_z$-up and $S_z$-down should completely coincide with each other.  It turns out that this is not the case, and a clear $S_z$ dependence of the EDCs was observed.  For instance, around the Fermi vector ($k_{\rm F}$) along the $\bar{\Gamma}$$\bar{K}$ direction corresponding to the FS angle $\phi$ of 0$^{\circ}$ (points A and B), the $S_z$-down EDC near $E_{\rm F}$ is clearly enhanced over to the $S_z$-up one around the peak top, although the difference between the two EDCs is much smaller than that in the IP polarization measurements [Fig. 1(d)].  Furthermore, the relative intensity of the near-$E_{\rm F}$ peak for the $S_z$-down and $S_z$-up spins is reversed along another $\bar{\Gamma}$$\bar{K}$ direction ($\phi$ = 60$^{\circ}$, which we call here the $\bar{\Gamma}$$\bar{K}'$ direction) as shown by the EDCs at point C.  This demonstrates a sign change of the OP spin polarization within the SS: Spins at the {\bf k}-space locations A and B point to the sample side, whereas those at location C point to the vacuum side.  Notably, there is no difference between $S_z$-up and $S_z$-down EDCs along the $\bar{\Gamma}$$\bar{M}$ direction (point D; $\phi$ = 30$^{\circ}$) within our experimental accuracy, suggesting that the sign change takes place along the $\bar{\Gamma}$$\bar{M}$ line. Also, one can see in Fig. 1(e) that there is no clear difference between $S_z$-up and $S_z$-down EDCs at {\bf k} near the $\bar{\Gamma}$ point (points E and F), suggesting that the $S_z$ component vanishes as the energy position on the SS approaches the Dirac point and the hexagonal warping weakens.

To discuss quantitatively the present observation, we plot in Figs. 1(f) and (g) the energy dependence of the IP ($P_{\phi}$) and OP ($P_z$) spin polarizations, respectively, for three representative {\bf k} points.  The spin polarization $P$ is defined as $P$ = (N$_{\uparrow}$-N$_{\downarrow}$)/(N$_{\uparrow}$+N$_{\downarrow}$) where N$_{\uparrow}$ and N$_{\downarrow}$ are the intensity of the up- and down-spin states, respectively.  One can clearly see that the $P_{\phi}$ value increases as $E_{\rm B}$ approaches $E_{\rm F}$, and it reaches $\sim$0.6 around $E_{\rm F}$.  In contrast, the magnitude of $|P_z|$ is much smaller, and at points A or C it reaches only $\sim$0.15, which is about 25\% of the maximum $P_{\phi}$ value.  We have confirmed that this characteristic behavior is reproducible by measuring several samples.  In addition, we have found that the exposure of an originally $p$-type Bi$_2$Te$_3$ samples to the helium gas in the vacuum chamber leads to the emergence of a similar snowflakelike FS due to the surface band bending \cite{HasanBi2Te3, KimPRB, Hoffmann} and that its spin texture is essentially the same as the bulk $n$-type samples. It is emphasized here that the clear observation of the OP spin component has become feasible by utilizing the high-flux and low-energy character of the xenon lamp \cite{SoumaRSIXe} as well as an efficient SR-ARPES spectrometer \cite{SoumaRSIspin}, which enable the high-energy or {\bf k} resolution measurement with higher statistics.

 \begin{figure}[b]
\includegraphics[width=3.4in]{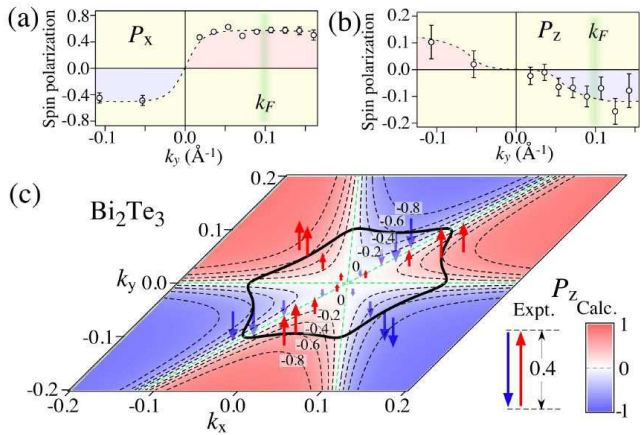}
\caption{(color online) (a),(b) Experimental spin polarization along the $x$ and $z$ axes, $P_x$ and $P_z$, of the SS along the $k_x$=0 cut, obtained by integrating the SR-ARPES intensity at $E_{\rm B}$ = 0.08-0.12 eV.  (c) Schematic picture of the $z$ component of the spin vector  as a function of {\bf k} in the SS of Bi$_2$Te$_3$ determined from the present ARPES experiment. The length of the arrow corresponds to the magnitude of $P_z$.  The data are symmetrized by taking into account the crystal symmetry. The intensity with gradual shading (red and blue) shows the calculated $P_z$ value based on the theory in Ref. \cite{FuPRL}.
}
\end{figure}

To analyze the {\bf k} dependence of the spin polarization in more detail, we plot in Figs. 2(a) and (b) the spin polarizations along the $x$- and $z$-axes measured in the $k_x$=0 cut [vertical solid (green) line in Fig. 1(b)] obtained by integrating the ARPES intensity at $E_{\rm B}$ = 0.08-0.12 eV; because of the rather large (40 meV) energy resolution of the spin-resolved measurements, the integration in this energy window gives a good representation of the spin polarization of the surface band as a function of {\bf k}.  One can see in Fig. 2(a) that the $P_x$ value at $k_y >$ 0 ($k_y <$ 0) is positive (negative), and its absolute value $|P_x|$ is nearly constant at $\sim$0.6 for $|k_y| \geq$ 0.03 ${\rm \AA^{-1}}$.  The sudden sign reversal in $P_x$ near $k_y$ = 0 is consistent with the IP spin helical texture.  We also found that the $z$-axis spin polarization $P_z$ shows a sign reversal along the same cut [Fig. 2(b)], as expected from the result in Fig. 1 where the $\bar{\Gamma}$$\bar{K}'$ direction at $\phi$ = 60$^{\circ}$ is identical to the $\bar{\Gamma}$$\bar{K}'$ direction at $\phi$ = 180$^{\circ}$ due to the three-fold rotational symmetry of the crystal.  However, the magnitude of $|P_z|$ stays nearly zero at $|k_y| \leq$ 0.03 ${\rm \AA^{-1}}$ and then gradually increases with $|k_y|$, reaching $\sim$0.15 near the $k_{\rm F}$ point.  This trend suggests that, unlike the behavior of $P_x$, $|P_z|$ grows systematically as the energy position on the SS moves away from the Dirac point at the zone center.

 To highlight the OP spin configuration obtained from the present experiment, we illustrate in Fig. 2(c) the magnitude of $P_z$ as a function of {\bf k}.  We also show numerically simulated $P_z$ values based on the $k\cdot p$ perturbation theory \cite{FuPRL} which reproduces the hexagonal deformation of the observed FS.  One can see that the measured $P_z$ switches the sign across the $\bar{\Gamma}$$\bar{M}$ line [straight dashed (green) lines] for every 60$^{\circ}$ step of $\phi$, and its absolute magnitude systematically increases as one moves away from the $\bar{\Gamma}$ point.  Such an overall experimental feature is in qualitative agreement with the numerical simulation.

  \begin{figure}[b]
\includegraphics[width=3.4in]{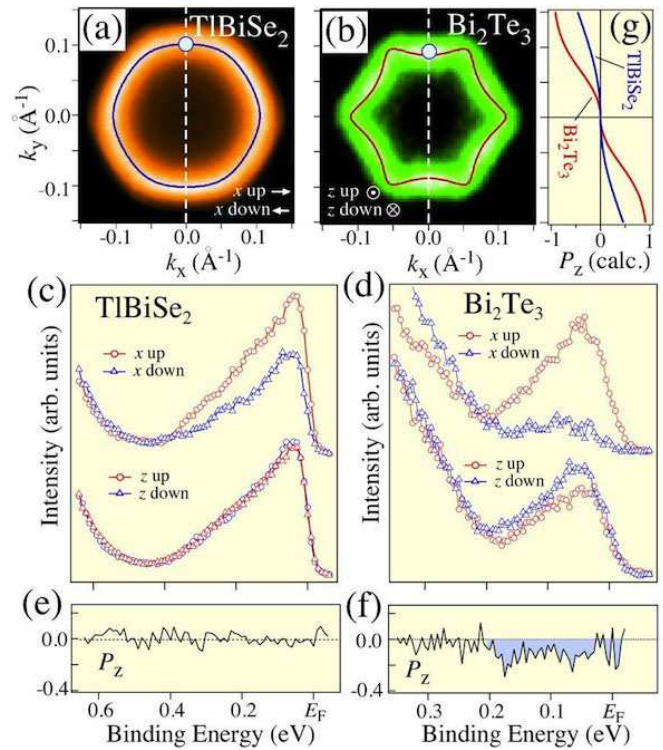}
\caption{(color online) (a),(b) Comparison of the FS between TlBiSe$_2$ and Bi$_2$Te$_3$.  Solid lines are the simulated FS using the theory in Ref. \cite{FuPRL}.  (c),(d)  SR-EDCs measured along the $\bar{\Gamma}$$\bar{K}$ line ($\phi$ = 0$^{\circ}$) for TlBiSe$_2$ and Bi$_2$Te$_3$, respectively.  (e),(f) Corresponding $E_{\rm B}$ dependences of $P_z$.  (g) Calculated $P_z$ value along the $k_x$=0 cut as a function of $k_y$ for TlBiSe$_2$ and Bi$_2$Te$_3$.
}
\end{figure}
 
 Next we discuss the relationship between the hexagonal warping of the FS and the OP spin polarization. We have performed SR-ARPES measurements on the recently discovered topological insulator TlBiSe$_2$ \cite{SatoPRL, HiroshimaPRL, StanfordPRL}.  As shown in Fig. 3(a), the FS of TlBiSe$_2$ is much closer to the circular shape as compared to the snowflakelike FS of Bi$_2$Te$_3$ [Fig. 3(b)], indicating a weaker warping effect in TlBiSe$_2$.  While an IP spin polarization is clearly observed [upper traces in Fig. 3(c)], the $S_z$-up and $S_z$-down EDCs of TlBiSe$_2$ overlap with each other [lower traces in Fig. 3(c)], indicating a negligible OP polarization which can also be seen in the $E_{\rm B}$ dependence of $P_z$ [Fig. 3(e)]. A side-by-side comparison of the Bi$_2$Te$_3$ data [Figs. 3(d) and (f)] vividly shows the difference between the two compounds in the magnitude of $P_z$.  In fact, the ratio of the OP spin component to the IP component in the $E_{\rm B}$ range of $E_{\rm F}$-0.2 eV is estimated to be 6$\pm$6$\%$ and 20$\pm$4$\%$ for TlBiSe$_2$ and Bi$_2$Te$_3$, respectively.  We have also confirmed the negligibly small $P_z$ value in TlBiSe$_2$ by measuring data at various {\bf k} points.  These results point to an intimate connection between the hexagonal warping of the FS and the magnitude of $P_z$ \cite{Pz}, although it is difficult to rule out all the other possibilities at the moment.  To further examine the present observation, we reproduced the observed FS shape of TlBiSe$_2$ and Bi$_2$Te$_3$ by using the theoretical band dispersion with the hexagonal warping term \cite{FuPRL} [see the solid curves in Figs. 3(a) and (b)] and calculated the $P_z$ value along the $k_x$=0 cut [dashed lines in Fig. 3(a)] as shown in Fig. 3(g).  Apparently, the calculated $P_z$ value in TlBiSe$_2$ is much smaller than that in Bi$_2$Te$_3$ ($\sim$0.2 and $\sim$0.6 at $k_{\rm F}$, respectively), which is qualitatively consistent with what we observed.

 However, the magnitude of $P_z$ is quantitatively different between theory and experiment, as seen in Fig. 3:  For instance, at the $k_{\rm F}$ point along the $\bar{\Gamma}$$\bar{K}$ direction ($\phi$ = 0$^{\circ}$) where $|P_z|$ becomes maximal on the FS, the theoretical $P_z$ of $\sim$0.6 [Fig. 3(g)] is $\sim$4 times larger than the experimental $P_z$ of $\sim$0.15 [Fig. 3(f)].  The smaller $P_z$ in the experiment cannot be attributed to extrinsic effects in the experiment such as the instrumental asymmetry of up- and down-spin channel or a nonlinearity in the detector sensitivity to the actual polarization, because the experimental value of $P_z$ is systematically reversed across $k_y$=0 with a similar magnitude relative to $k_y$=0  [see Fig. 2(b)] and the experimental scattering asymmetry was confirmed to be proportional to the theoretical spin polarization by measuring  Au(111) as a reference.  Therefore, there must be an intrinsic reason for the weakening of $P_z$.  One possible explanation is the many-body effects that lead to an electronic order like the spin density wave where the FS is nested with the wave vector connecting parallel segments of the hexagonally warped FS \cite{FuPRL, HiroshimaBi2Se3}.  Another possibility is the presence of an interband scattering channel \cite{KimPRB} that degrades the absolute surface spin polarization through the scattering between the surface band and the non-spin-polarized bulk band. It is desirable that a concrete theoretical model is developed to explain the unexpectedly small OP polarization.
 
  The presence of the OP spin polarization provides a novel mechanism for the ``mass acquisition'' of the Dirac particles on the surface \cite{FuPRL}, which is a prerequisite of exotic topological phenomena like the half-integer quantum Hall effect and the quantized magnetoelectric effect \cite{QiPRB, QiScience}.  When the SS has a purely IP helical spin texture, the gapping of the Dirac cone requires an OP magnetic field $B_{\perp}$ or magnetization $M_{\perp}$.  On the other hand, when the SS has a finite OP spin component $S_z$, application of an IP magnetic field $B_{\parallel}$ would also lead to a gap opening due to the coupling between $B_{\parallel}$ and $S_z$ \cite{FuPRL}.  This way of gapping would be useful for achieving a massive Dirac state since it can essentially avoid the complicated orbital effects which generally couple to the IP {\bf k}'s and spins \cite{SCZhangPRB}.  Hence our observation of the OP spin component assures a promising direction for realizing novel quantum state and device applications that require a gapped SS.

 In conclusion, our SR-ARPES measurements of Bi$_2$Te$_3$ provide direct and compelling evidence for the OP spin component on the gapless SS, by simultaneously determining the IP and OP spin textures at various {\bf k} points in the Brillouin zone.  Moreover, we found that a stronger hexagonal deformation of the FS leads to a larger OP spin component, by comparing the observed FS and the SR-ARPES data with those of TlBiSe$_2$.  Intriguingly, the OP polarization is found to be much smaller than that expected from existing theory in both Bi$_2$Te$_3$ and TlBiSe$_2$, which poses an interesting question for future studies.

\begin{acknowledgments}
We thank T. Arakane for the development of an {\it in-situ} sample rotation system for the SR-ARPES measurements and Zhi Ren for the preliminary crystal growth of Bi$_2$Te$_3$.  This work was supported by JSPS (KAKENHI 19674002 and the Next-Generation World-Leading Researchers Program), JST-CREST, MEXT of Japan (Innovative Area ``Topological Quantum Phenomena''), and AFOSR (AOARD 10-4103).
\end{acknowledgments}

{\it Note added.} Recently, we became aware of a related SR-ARPES study on more highly electron-doped Bi$_2$Te$_3$ \cite{XuSpin23}. The reported $P_z$ value (0.3) is larger than that of the present study (0.15). This difference is consistent with the $k {\cdot} p$ theory \cite{FuPRL} and is attributed to the difference in the magnitude of the warping factor $w$ \cite{XuSpin23} between the two cases (1.1 in the present sample, as opposed to 1.6 in Ref. \cite{XuSpin23}).


\begin{references}

\bibitem{HasanRMP}
M. Z. Hasan and C. L. Kane, Rev. Mod. Phys. {\bf 82}, 3045 (2010).
\bibitem{HasanBi2Se3}
Y. Xia {\it et al.}, Nature Phys. {\bf 5}, 398 (2009).
\bibitem{YazdaniSTM}
P. Roushan {\it et al.}, Nature (London) {\bf460}, 1106 (2009).
\bibitem{YazdaniSb} 
J. Seo {\it et al.}, Nature (London) {\bf466}, 343 (2010).
\bibitem{AlpichshevPRL}
Z. Alpichshev {\it et al.}, Phys. Rev. Lett. {\bf104}, 016401 (2010).
\bibitem{FuPRL}
L. Fu, Phys. Rev. Lett. {\bf103}, 266801 (2009).
\bibitem{SCZhangPRB}
H.-J. Zhang {\it et al.}, Phys. Rev. B {\bf80}, 085307 (2009).
\bibitem{Louie}
O. V. Yazyev, J. E. Moore, and S. G. Louie, Phys. Rev. Lett. {\bf105}, 266806 (2010).
\bibitem{ShenBi2Te3}  
Y. L. Chen {\it et al.}, Science {\bf325}, 178 (2009).
\bibitem{SoumaRSIspin} 
S. Souma {\it et al.}, Rev. Sci. Instrum. {\bf81}, 095101 (2010).
\bibitem{SoumaRSIXe}
S. Souma {\it et al.}, Rev. Sci. Instrum. {\bf78}, 123104 (2007).
\bibitem{HasanBi2Te3}
D. Hsieh {\it et al.}, Nature (London) {\bf460}, 1101 (2009).
\bibitem{KimPRB}
S. R. Park {\it et al.}, Phys. Rev. B {\bf81}, 041405R (2010).
\bibitem{Hoffmann}
M. Bianchi {\it et al.}, Nat. Commun. {\bf1} : 128 (2010).
\bibitem{SatoPRL}
T. Sato {\it et al.}, Phys. Rev. Lett. {\bf105}, 136802 (2010).
\bibitem{HiroshimaPRL}
K. Kuroda {\it et al.}, Phys. Rev. Lett. {\bf105}, 146801 (2010).
\bibitem{StanfordPRL}
Y. L. Chen {\it et al.}, Phys. Rev. Lett. {\bf105}, 266401 (2010).
\bibitem{Pz}
The close connection between the hexagonal warping and the magnitude of $P_z$ may also be recognized by looking at the data of Bi$_2$Te$_3$ in Fig. 2(b) where the $P_z$ value is reduced near the $\bar{\Gamma}$ point as the hexagonal warping becomes weaker near the Dirac point \cite{ShenBi2Te3,HiroshimaBi2Se3}.
\bibitem{HiroshimaBi2Se3}
K. Kuroda {\it et al.}, Phys. Rev. Lett. {\bf105}, 076802 (2010)
\bibitem{QiPRB}
X. L. Qi, T. L. Hughes, and S. C. Zhang, Phys. Rev. B {\bf78}, 195424 (2008).
\bibitem{QiScience}
X.-L. Qi {\it et al.}, Science {\bf323}, 1184 (2009).
\bibitem{XuSpin23}
S.-Y. Xu {\it et al.},  arXiv:1101.3985.

\end{references}
\end{document}